**Gendered Pixels: Exploring Gender Differences in Computer-Mediated Self-Presentation among Douyu Live Streamers**

Mengxiao Zhu[1,2], Xiyue Wang[3,*], Chunke Su[4], Han Zhao[5], Cuihua Shen[6]

[1] Department of Communication of Science and Technology, University of Science and Technology of China, Hefei, China

[2] Anhui Province Key Laboratory of Science Education and Communication, University of Science and Technology of China, Hefei, China

[3] Department of Communication Studies, University of Kansas, Lawrence, USA

[4] Department of Communication, University of Texas, Arlington, USA

[5] Institute of Advanced Technology, University of Science and Technology of China, Hefei, China

[6] Department of Communication, University of California, Davis, USA

*Corresponding author: Xiyue Wang

Department of Communication Studies

University of Kansas

Bailey Hall, 1440 Jayhawk Blvd.

Lawrence, KS 66045, USA

E-mail address: xiyue.wang@ku.edu

## Abstract

Live streaming platforms, as computer-mediated communication (CMC) systems, provide streamers with a range of tools, such as webcams, beauty filters, and stream titles, to shape their online personas in ways that either conform to or deviate from viewers' expectations. Drawing on gender role and CMC theories, this study examines how streamers leverage CMC self-presentation tools to fulfill gender role expectations and their associated live streaming outcomes. Analyzing data collected from 867 streamers and 94,227 streams on Douyu, a popular Chinese live streaming platform, we find that although both female and male streamers make extensive use of CMC tools, female streamers are more likely to employ visual strategies, such as webcams and beauty filters, than male streamers. We further find that although different tools have varying associations on streamers' earnings and audience engagement, the benefits of webcam use are weaker for female than for male streamers. These findings underscore the complex interplay between gender roles and technology use in the live streaming domain.

Live streaming has rapidly emerged as a popular form of online interaction worldwide. The global live streaming market grew from $1.24 billion in 2022 to $1.49 billion in 2023, representing an annual growth rate of 20.6%, and is projected to reach $3.21 billion by 2027 (Research and Markets, 2024). Live streaming provides individuals from diverse backgrounds with opportunities to express themselves, showcase their talents, and reach niche audiences. While live streaming offers a platform for inclusivity and diversity, gender gaps, which have been widely documented across various social domains, are also evident in this context (Xu et al., 2022). For example, gaming live streaming, as a dominant category on major platforms, remains heavily male-dominated (Sjöblom et al., 2019), with women accounting for only a small proportion of top-performing streamers (Ceci, 2024). To uncover the mechanisms underlying these gender gaps, researchers have examined the distinct strategies that male and female streamers use to engage viewers (e.g., Zhang & Hjorth, 2019; Anciones-Anguita & Checa-Romero, 2024). Despite these insights, previous research has largely overlooked gendered strategies involving the use of live streaming tools.

As computer-mediated communication (CMC) systems, live streaming platforms provide streamers with a range of tools, such as webcams, beauty filters, and stream titles (Deng et al., 2021; Sjöblom et al., 2019). These tools enable streamers to curate their online personas in ways that either conform to or deviate from viewers' expectations, potentially leading to different interactional and economic outcomes. To examine how streamers of different genders strategically use CMC tools and how such tool use is associated with streaming outcomes, we draw on gender role theory (Eagly & Karau, 2002) and relevant CMC theories (Walther, 2011). Specifically, we examine whether streamers' use of CMC tools is associated with better outcomes, including chat message volume and earnings, and whether male and female streamers differ in their patterns of tool use. Additionally, this study explores whether gender moderates the relationship between CMC tool use and live streaming outcomes.

By analyzing data collected from 867 streamers and their 94,227 streams across the three most-viewed channels on Douyu, a popular live streaming platform in China, this study contributes to CMC research and gender studies in several ways. First, it integrates gender role theory with CMC theories to examine streamers' behaviors and outcomes, allowing for a more comprehensive analysis of how gender roles are expressed and negotiated in digital interactions. Second, by collecting and analyzing large-scale behavioral data from a prominent Chinese live streaming platform, this study complements existing research on gender gaps in live streaming, which has often relied on qualitative approaches and focused primarily on Western contexts. Finally, this study highlights gender gaps in streamers' strategic use of CMC tools for self-presentation and outcome achievement, thereby enriching our practical understanding of how technologies shape and constrain gendered behaviors within the live streaming domain.

## Literature Review

### *Live Streaming and Gender Differences*

Live streaming is an Internet-based interactive medium through which streamers broadcast real-time audio and video content to interested viewers (Deng et al., 2021). Streamers produce a wide range of content, including video game playthroughs, educational lectures, and live music performances. Besides watching streamed content, viewers can interact with streamers and other viewers by sending chat messages and virtual gifts (Li & Guo, 2021). Revenue from virtual gifts is typically shared between the streamer and the platform, making virtual gifting a primary source of income for streamers (Johnson & Woodcock, 2019). Because chat messages and virtual gifts are highly visible and measurable indicators of audience engagement and financial support, they have become two of the most commonly used metrics for evaluating streamer success (Choi et al., 2022; Li & Guo, 2021; Liu et al., 2022).

As offline gender inequalities often extend into online spaces, recent research has examined gendered performances (Zhang & Hjorth, 2019), gender-based harassment (Nakandala et al., 2017), and the role of live streaming community guidelines and platform policies in shaping gender inequality (Ruberg, 2021). For instance, prior research showed that female streamers receive more chat messages and virtual gifts than male streamers, which may be attributed to female streamers' higher levels of interactivity and audience engagement, as well as the relative scarcity of female streamers on live streaming platforms (Choi et al., 2022; Wang & Li, 2020; Wolff & Shen, 2022). Meanwhile, gendered advantages in audience engagement do not necessarily translate into equal recognition or treatment. Compared with male streamers, female streamers are often perceived as less competitive and less professional and receive more objectifying comments, particularly in game live streaming contexts (Ruvalcaba et al., 2018).

Despite growing attention to gender differences in live streaming, one underexplored area concerns how streamers use the CMC tools provided by live streaming platforms and whether such tool use differs by gender. As CMC systems, live streaming platforms offer a variety of technical tools, such as webcams, beauty filters, and customizable stream titles, that allow streamers to manage their self-presentation, enhance communication with viewers, and potentially gain audience support (Deng et al., 2021; Sjöblom et al., 2019). However, the extent to which these CMC tools are associated with improved streaming outcomes, as well as whether male and female streamers use these tools differently, has not been empirically investigated. To address this gap, this study integrates gender role theory (Eagly & Karau, 2002) with relevant CMC theories including social presence theory, hyperpersonal theory, and social information processing theory (Walther, 2011) to examine streamers' use of CMC tools and gender differences in their usage.

***Visual Presence through Webcams***

One main feature of live streaming is its capacity to create a sense of visual presence for viewers through the use of webcams (Mueser & Vlachos, 2018). According to social presence theory, media that convey more communication cues tend to generate a stronger sense of warmth, involvement, and interpersonal connection (Walther, 2011). When streamers turn on their webcams to show their appearance and real-time physical surroundings, they provide viewers with richer nonverbal cues, such as eye contact, facial expressions, and bodily gestures. These visual cues can enhance viewers' perceptions of the streamer as authentic and socially present (Gunawardena & Zittle, 1997; Yim et al., 2017). Increased social presence can also foster trust and reduce the psychological distance between streamers and viewers (Jiang et al., 2019; Zhang et al., 2021), potentially encouraging viewers to engage through chat messages and virtual gifts. Furthermore, webcams allow viewers to observe streamers' immediate reactions to chats and gifts, fostering a sense of engagement and immediacy (Guarriello, 2019), and encouraging further viewer contributions (Lin, 2021). Therefore, we propose the following hypothesis:

**H1: Streamers with their webcam on are associated with better live streaming outcomes, including chat messages and earnings.**

While webcam use may benefit streamers in general, streamers may use webcams differently in response to gendered expectations to avoid online bias or discrimination (Eagly & Karau, 2002). As gender role theory suggests (Eagly & Karau, 2002), men are often expected to display agentic qualities, such as assertiveness, strength, independence, and emotional restraint, whereas women are typically expected to display communal qualities, such as nurturance, empathy, accommodation, and emotional expressiveness. In live streaming contexts, these gender expectations may translate into distinct presentation norms. Male streamers are often encouraged to showcase strength and expertise, whereas female streamers may face stronger expectations to emphasize visual attractiveness and relational affinity (Manago et al., 2008). Consequently, female streamers may feel more compelled to use webcams to maintain a higher social presence through visual engagement and emotional expression (Manago et al., 2008). In contrast, male streamers are expected to demonstrate agentic qualities, such as gaming skill or technical expertise, without relying as heavily on webcam use. Therefore, we propose the following hypothesis:

**H2: Female streamers are more likely to turn on webcams than male streamers.**

Besides the above hypotheses, this study also explores whether gender moderates the association between webcam use and live streaming outcomes. Gender may shape the benefits of webcam use in two competing ways. On the one hand, webcam use may produce stronger positive effects for female streamers because it allows them to conform to established expectations of femininity, such as being visually engaging, emotionally expressive, and relationally responsive. This conformity may elicit greater viewer engagement and support, including more chat messages and virtual gifts. Prior research suggests that women have advantages in using nonverbal communication to facilitate interpersonal interaction (Gunnery et al., 2013). In live streaming, female streamers may therefore be especially able to use visual and nonverbal cues to enhance viewers' experience and generate positive responses.

On the other hand, because female streamers may already be expected to use webcams by default, webcam use may not necessarily generate additional rewards for them. Instead, female streamers who do not use webcams may be penalized for violating viewers' gendered expectations. In contrast, male streamers experience less pressure to use webcams because expectations for them often focus on skills, knowledge, humor and performance (Eagly et al., 1995). As a result, male streamers who turn on their webcams may exceed viewers' expectations and receive additional engagement or support. Given these competing possibilities, we propose the following research question:

**RQ1: How does gender moderate the association between webcam use and live streaming outcomes, including chat messages and earnings?**

***Visual Appeal through Beauty Filters***

Another popular set of tools offered by live streaming platforms are beauty filters, which allow users to edit, refine, and enhance their appearance in photos and real-time (Lian & Chen, 2023; Elias & Gill, 2018). Research has examined beauty filter use on social media and dating apps, with particular attention to users' motivations for using beauty filters (e.g., Herring et al., 2024) and their potential negative consequences, such as facial and body dissatisfaction (e.g., Tiggemann, 2022). However, the use and effects of beauty filters in the context of live streaming remain underexplored.

This study investigates the use and effects of beauty filters on live streaming platforms through the perspective of hyperpersonal models of CMC (Walther, 1996). The hyperpersonal model suggests that users can strategically use CMC tools to engage in selective self-presentation by emphasizing the most appealing and desirable aspect of themselves while concealing less favorable aspects (Walther, 1996). Prior research has shown that physical attractiveness can generate positive professional outcomes, including improved sales performance, greater perceived trustworthiness, and higher customer satisfaction ratings (Söderlund & Julander, 2009; Talamas et al., 2016).

On CMC platforms, where users often have limited access to non-verbal cues, they tend to form faster, potentially more stereotypical impressions of others (Walther & Parks, 2002). In live streaming where content sharing is highly visual, the importance of visual appeal may be especially pronounced. Moreover, as suggested by the hyperpersonal model, receivers tend to exaggerate their perception of a sender based on limited initial cues (Walther, 1996). By enhancing streamers' visual appearance, beauty filters may increase viewers' perceptions of streamers as appealing, thereby encouraging viewer engagement and support. Therefore, we propose the following hypothesis:

**H3: Streamers using beauty filters are associated with better live streaming outcomes, including chat messages and earnings.**

Although both female and male streamers may benefit from using beauty filters, they may differ in their likelihood of using them. In general, women often face greater pressure than men to conform to idealized beauty standards and fulfill societal expectations of femininity (Strahan et al., 2006). Such pressures may increase women's self-monitoring of their appearance, motivate them to maintain an attractive look (Grabe & Hyde, 2009) and encourage greater engagement in photo-enhancement behaviors and profile management (Haferkamp et al., 2012). On live streaming platforms, beauty filters provide female streamers with a readily available tool for enhancing their appearances in ways that align with these gendered expectations. Conversely, male streamers may face less societal pressure to prioritize visual attractiveness and therefore may have less social motivation to use beauty filters (Benze & Declercq, 1985). Thus, we propose the following hypothesis:

**H4: Female streamers are more likely to use beauty filters than male streamers.**

Building on these hypotheses, we further posit that gender may moderate the association between beauty filter use and live streaming outcomes in two competing directions. On the one hand, beauty filter use may benefit female streamers more than male streamers because enhanced physical attractiveness is more closely aligned with gender role expectations for women (Travis et al., 2000) than for men (Pleck et al., 1993). In this sense, beauty filters may help female streamers conform to viewers' expectations of femininity, potentially increasing viewer engagement and support. For male streamers, however, beauty filter use may be less consistent with conventional masculine norms. Because beauty filters and makeup are often viewed as diverging from typical masculine practices, male streamers who use beauty filters may face negative audience reactions, such as reduced viewer engagement.

On the other hand, the positive effect of beauty filter use may be weaker for female streamers than for male streamers if beauty filters undermine female streamers' perceived credibility, particularly among male viewers. For instance, McGloin & Denes (2018) found that men perceived a woman using an idealized image as less trustworthy than the same woman using a less enhanced image. Given that live streaming platforms are often dominated by male viewers (Sjöblom et al., 2019), female streamers who use beauty filters are more likely to be perceived as less authentic or trustworthy, which could weaken or even reduce the benefits of beauty filter use for live streaming outcomes. Considering these competing mechanisms, we propose the following research question:

**RQ2: How does gender moderate the association between beauty filter use and live streaming outcomes, including chat messages and earnings?**

***Tagline Attraction through Stream Titles***

Stream titles are brief descriptions or taglines that streamers use to label their streams and attract potential viewers. These titles provide viewers with an initial impression of the stream content and the streamer's identity before they enter the live stream. As suggested by social information processing theory of CMC, when nonverbal cues are limited or unavailable, communicators adapt by relying more heavily on available verbal and linguistic cues to form impressions and develop relationships in CMC environments (Walther, 1992). On live streaming platforms, viewers typically encounter textual and visual information about a stream before deciding whether to enter it. As a result, stream titles serve as an important pre-stream communication tool through which streamers can strategically signal content, personality, and interactional style to potential viewers (Sjöblom et al., 2019).

Because stream titles shape viewers' initial expectations, titles that align with gender role expectations may generate greater viewer interest and engagement. For example, female streamers may use titles that emphasize collaboration or community, whereas male streamers may use titles that highlight competition or gaming mastery. By using titles that conform to viewers' gendered expectations, streamers may create a sense of familiarity and expectation consistency, thereby encouraging viewers to enter the stream, participate in chat, and offer financial support. Therefore, we propose the following hypothesis:

**H5: Streamers whose stream titles conform better to gender role expectations are associated with better live streaming outcomes, including chat messages and earnings.**

To conform stream titles to gender role expectations, male and female streamers may rely on different linguistic strategies. Prior studies found gender differences in language use on social media. Women tend to use more inclusive language, such as pronouns like "we" and "you," as well as more emotional terms and informal elements such as emoticons and abbreviations. Men, by contrast, tend to use more language associated with dominance, action, numbers and swearing (Bamman et al., 2014). Similar gendered language patterns may appear in live streaming titles, as streamers strategically present themselves in ways that align with broader expectations of masculinity and femininity. Specifically, male streamers may be more likely to emphasize productivity and achievements, whereas female streamers may emphasize interpersonal relationships and emotional connection (Wang & Li, 2020). Therefore, we propose the following hypothesis:

**H6: Female and male streamers tend to customize their stream titles in different ways to conform to their gender role expectations.**

Building on these hypotheses, we further examine whether gender moderates the association between gender-role-conforming stream titles and live streaming outcomes. Stream titles that align with gender role expectations may not benefit male and female streamers equally because viewers may apply different evaluative standards depending on the streamer's gender. However, existing research provides limited evidence on the direction and strength of this moderating effect in live streaming contexts. Therefore, we propose the following research question:

**RQ3: How does gender moderate the association between stream title conformity to gender role expectations and live streaming outcomes, including chat messages and earnings?**

## Methods

### *Douyu: The Sample Live Streaming Platform*

Data for this study were collected from Douyu, one of China's largest live streaming platforms. Launched in 2014, Douyu has approximately 50,000 streamers and 3 million active viewers (Bojiang, 2024), and is often described as China's counterpart to Twitch (Dredge, 2020). Douyu organizes streams into channels based on content. These include channels dedicated to specific games, such as League of Legends and Minecraft, as well as channels featuring nongaming content, such as food, outdoor activities, music, and lifestyle content. Similar to those of other live streaming platforms, Douyu's audience is predominantly young and male: approximately 93% of its viewers are men, and 61% are between 20 and 25 years old (DouyuTV, 2021).

As shown in Figure 1, Douyu provides streamers with various CMC tools, including webcams, beauty filters, customizable stream titles, and cover images. By default, the webcam and a beauty filter are activated at the beginning of a stream, although streamers may choose to disable either feature. Figure 2 presents the viewer interface, where viewers can interact with streamers and one another by sending chat messages and virtual gifts. Virtual gifts range in price from 0.1 to 5,000 Chinese yuan (approximately 0.01 to 700 US Dollars). These gifts appear in the chat area and are accompanied by elaborate animations, such as rockets or spacecraft. Revenue from virtual gifts represents a primary source of income for streamers and an important source of revenue for Douyu[1]. Streamers can also increase their level by completing daily streaming tasks, such as meeting duration requirements, and by receiving virtual gifts. Streamer level therefore serves as an indicator of a streamer's activity, engagement, and popularity on the platform.

[1] Douyu has not disclosed the specific percentage cut from streamers' gifts.

**Figure 1**

*Screenshot of Stream Options on Douyu's Mobile App from a Streamer's Perspective*

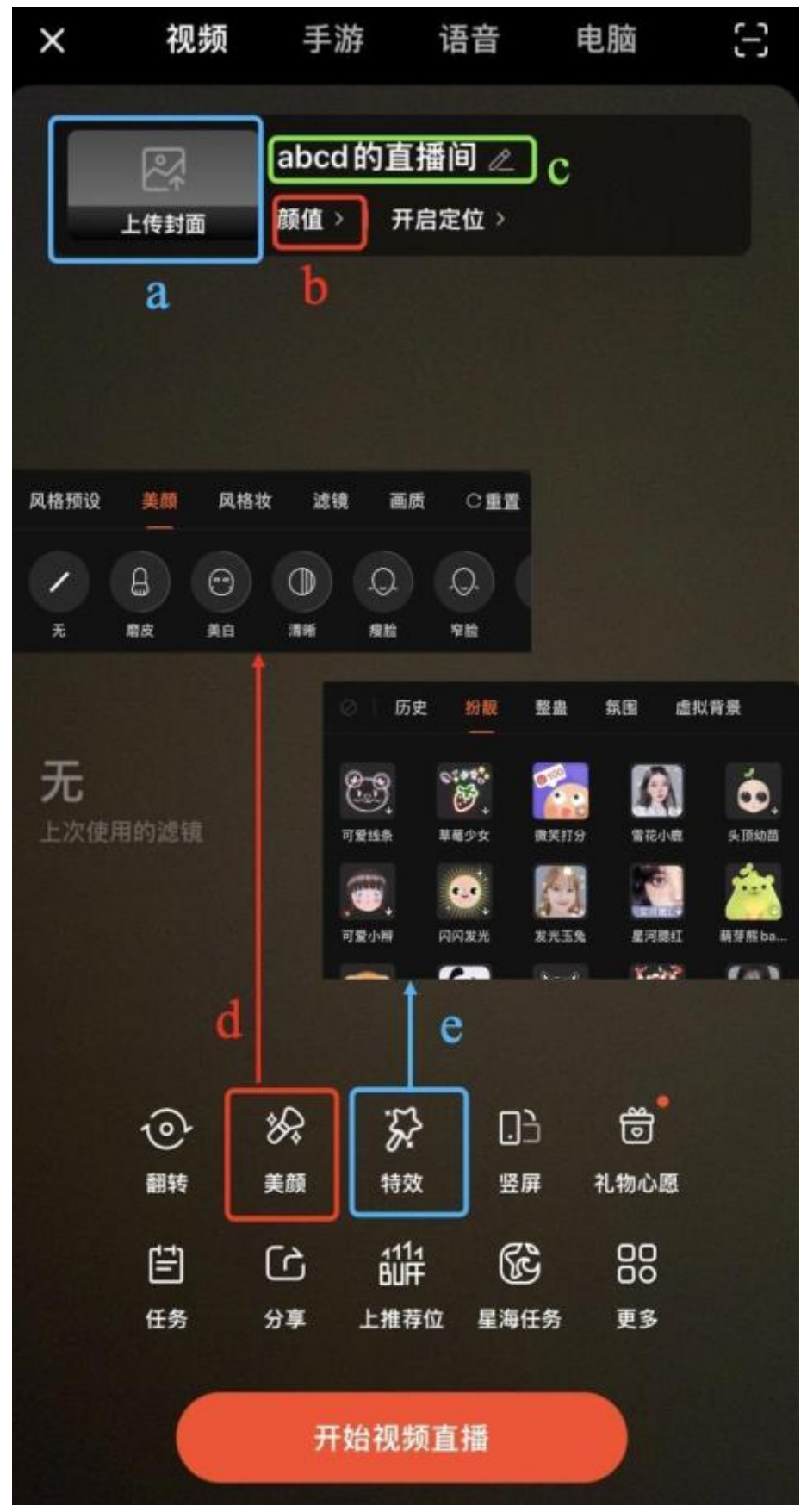

*Note.* a. Stream cover image; b. Stream channel; c. Stream title; d. Beauty filters (facial retouching, filters, and makeup); e. Special effects (stickers with special visual effects)

**Figure 2**

*Screenshot of a Stream in Progress on Douyu from a Viewer's Perspective*

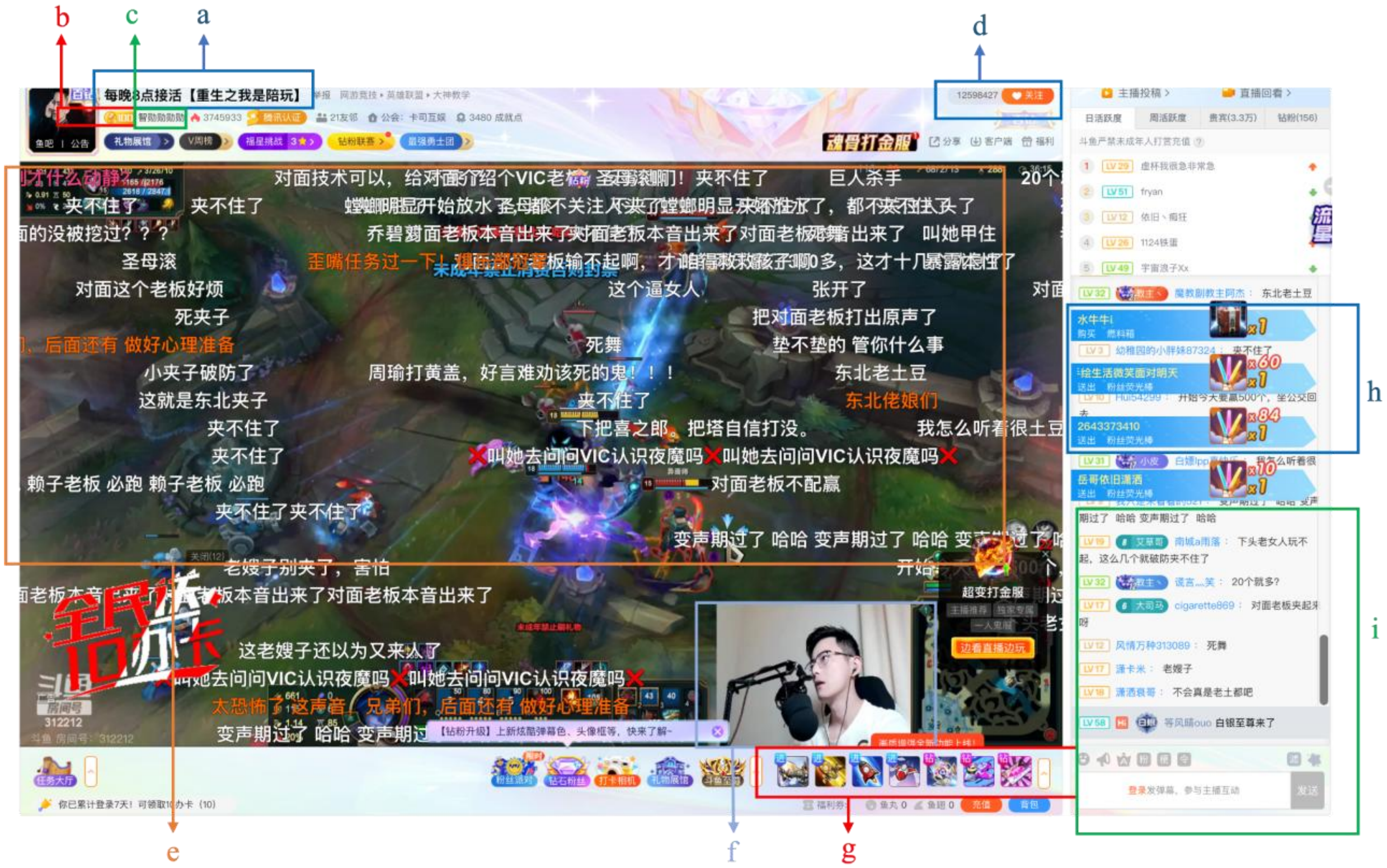

*Note.* a. Stream title; b. Streamer level; c. Streamer username; d. Number of followers; e. Chat messages (danmaku; real-time viewer messages displayed over the stream); f. Webcam display; g. Virtual gift option; h. Virtual gift animation; i. Chat box

***Data Collection***

We collected data from the three most-viewed channels on Douyu (DouyuTV, 2021), including League of Legends (LOL), Yanzhi, and Honor of Kings (HOK). LOL and HOK are gaming channels named after the games featured in their streams, whereas Yanzhi is a more generally themed entertainment channel. Both LOL and HOK are popular multiplayer online battle arena games, but they are designed for different devices: LOL is played on personal computers, whereas HOK is primarily played on mobile devices. These technological differences may shape streamers' broadcasting practices. LOL streamers generally broadcast from personal computers and can readily use webcams. HOK streamers may broadcast directly from mobile devices or mirror their screens to personal computers; webcam use is available only when streaming through a computer. Unlike the game-focused LOL and HOK channels, the Yanzhi channel features a wide range of content, including singing, dancing, casual conversation, and lifestyle vlogging. Streamers in this channel are predominantly female (Xu et al., 2022), and frequently attract viewers by highlighting personal characteristics such as physical appearance and vocal qualities (Liu et al., 2022).

We first gathered a list of all active streamers who conducted at least one live streaming session, hereafter referred to as a "stream," in one of the three channels between April 3 and April 9, 2022. We then applied the following inclusion criteria: a) the streamer's gender could be identified from available visual or audio cues; b) the account was operated by an individual rather than an organization; c) the stream was conducted in Chinese; d) the chat box was accessible to all viewers; e) each stream lasted longer than 10 minutes; and f) the streamer's average weekly popularity exceeded the mean popularity of all streamers in the same channel. A total of 1,454 streamers met these criteria, including 508 in LOL, 457 in HOK, and 489 in Yanzhi.

We subsequently collected data on the streaming activities of these 1,454 streamers from June 6 to July 31, 2022. Using an application programming interface, we captured textual and behavioral data from their streams. We then cleaned the data to distinguish live sessions from sessions in which previously recorded content was replayed. Only live sessions were retained; the detailed identification procedure is reported in Appendix 1. To focus on consistently active streamers, we also excluded streamers who broadcast less than once per week during the study period.

The final dataset included 867 streamers (including 339 in LOL, 301 in HOK, and 227 in Yanzhi) with an average of 279.16 streaming hours (*Median* = 266.33, *SD* = 146.08), and their 94,227 streams attracted an average of 71,534.86 viewers (*Median* = 22,806, *SD* = 120,829.98). The dataset contained streamer attributes, including streamer ID and level, and stream records, including the stream title, channel tag, date and time of the stream, viewer count, virtual gift count, popularity value, chat messages, and platform notifications. Streamer IDs were used to link the two types of data.

### *Measures*

**Gender.** Streamer gender was coded as a binary variable based on visual and audio cues available in the stream content (male = 0, female = 1). Two independent coders assessed each streamer and reached a 100% agreement. As shown in Table 1, 53.2% of the sampled streamers were coded as women and 46.8% were coded as men. This gender composition should not be interpreted as representative of the overall gender distribution of streamers on Douyu.

**Chat messages.** Viewer engagement was measured using the cumulative number of chat messages received by each streamer across all streams conducted between June 6 and July 31, 2022. During the study period, streamers received an average of 144,413.59 chat messages (*Median* = 24,971, *SD* = 648,691.09).

**Earnings.** Streamer earnings were calculated by summing the monetary value of all virtual gifts received during the study period. On average, streamers earned 170,498.01 Chinese Yuan (approximately $23,647 US Dollars; *Median* = 47,935.16, *SD* = 513,001.03).

**Webcam use.** Webcam use was manually coded as a binary variable indicating whether the streamer used a webcam during the observed streams (yes = 1, no = 0). Two independent coders reviewed the stream content (Cohen's Kappa = 0.96). Overall, 55% of the streamers used a webcam, whereas 45% did not.

**Beauty filter use.** Beauty filter use was coded as a binary variable. Streamers who used at least one beauty filter were coded as 1, whereas those who did not use a beauty filter were coded as 0. Because beauty filters modify webcam images, this measure was applicable only to streamers who used webcams. Two independent coders reviewed the stream videos (Cohen's Kappa = 0.96).

**Stream title customization.** Topic modeling was conducted to identify how streamers customized their stream titles. Douyu stream titles are limited to 20 Chinese characters. Before conducting the topic analysis, we removed stop words and special characters, including punctuation marks and emojis. We then segmented the Chinese text using the *jieba* word-segmentation tool (Day & Lee, 2016). The automatically segmented text was manually reviewed and adjusted to improve segmentation accuracy.

Next, we applied the BERTopic model (Grootendorst, 2022), a topic-modeling approach that identifies semantically similar texts using BERT-based embeddings. Each topic was represented by a set of up to 10 core words ranked by their frequencies. Two coders reviewed the topics, their representative keywords, and the original stream titles using an open-coding approach. Through this process, four overarching themes were identified: self-expression, impression management, interaction, and stream content. The two coders then independently assigned each BERTopic-generated topic to one of these four themes (Cohen's Kappa = 0.78).

**Title conformity to gender role expectations.** We operationalized a stream title's conformity to gender role expectations as its linguistic similarity to titles used by other streamers in the same gender group. We first used the CountVectorizer function in the scikit-learn Python library to convert the stream titles into numerical vectors (Pedregosa et al., 2011). We then grouped the title vectors according to streamer gender and calculated the centroid, or mean vector, for each gender group. For each title, we calculated its Euclidean distance from the centroid of the streamer's respective gender group. We multiplied this distance by –1 so that higher values indicated greater similarity to the linguistic patterns of the corresponding gender group. Thus, higher scores represented greater conformity to gender-typical stream title conventions, whereas lower scores represented greater deviation from those conventions. For interaction models, title conformity was mean-centered prior to constructing interaction terms.

***Control Variables.*** We controlled streamer level, channel, viewer count, and total streaming duration. Streamer level and channel were obtained directly from the platform. Viewer count was measured as the total number of viewers who watched a streamer's streams during the study period. Total streaming duration was calculated by summing the duration, in hours, of all live streams conducted by each streamer during the study period.

To correct the skewness of the count variables in our analysis including chat messages, earnings, and viewer counts, we applied a logarithmic transformation, which is a common practice in treating skewed variables (Premkumar & Bhattacherjee, 2008).

**Table 1**

*Descriptive Statistics of Streamers*

| Variable | Total ($N$ = 867) | | | Male ($n$ = 406, 46.8%) | | | Female ($n$ = 461, 53.2%) | | |
|---|---|---|---|---|---|---|---|---|---|
| | Mean | Median | S.D. | Mean | Median | S.D. | Mean | Median | S.D. |
| Webcam | 0.55 | — | — | 0.27 | — | — | 0.80 | — | — |
| Beauty filter | 0.84 ($N$ = 479) | — | — | 0.32 ($n$ = 111) | — | — | 0.99 ($n$ = 368) | — | — |
| Title conformity | -2.71 | -2.64 | 0.63 | -2.80 | -2.78 | 0.61 | -2.64 | -2.55 | 0.64 |
| Streamer level | 49.49 | 46.00 | 19.23 | 50.98 | 48.00 | 20.30 | 48.18 | 45.00 | 18.16 |
| Duration | 279.16 | 266.33 | 146.08 | 305.22 | 291.42 | 170.45 | 256.21 | 245.00 | 116.06 |
| Chat messages | 144,413.59 | 24,971.00 | 648,691.09 | 248,095.82 | 30,163.00 | 873,344.18 | 53,101.26 | 23,314.00 | 320,661.49 |
| Chat messages (log) | 10.39 | 10.13 | 1.27 | 10.69 | 10.31 | 1.57 | 10.13 | 10.06 | 0.84 |
| Earnings | 170,498.01 | 47,935.16 | 513,001.03 | 208,695.49 | 35,636.94 | 656,149.13 | 136,857.72 | 57,073.19 | 337,593.34 |
| Earnings (log) | 10.70 | 10.78 | 1.71 | 10.44 | 10.48 | 2.03 | 10.92 | 10.95 | 1.33 |
| Viewer counts | 71,534.86 | 22,806.00 | 120,829.98 | 93,954.36 | 44,670.00 | 116,819.36 | 51,790.15 | 18,276.00 | 120,981.20 |

### *Analysis*

We first conducted *t*-tests to examine the gender differences in streamers' chat messages and earnings. To explore the associations of webcam use (H1), beauty filter use (H3) and title conformity to gender role expectation (H5) with chat messages and earnings, we estimated a series of linear regression models. We also conducted chi-square tests to examine whether webcam use (H2) and beauty filter use (H4) differed between male and female streamers. To examine whether male and female streamers customized their stream titles differently (H6), we compared the distribution of the identified title themes across gender groups. Finally, we estimated additional regression models that included interaction terms between gender and webcam use (RQ1), beauty filter use (RQ2), and stream title conformity to gender role expectations (RQ3).

## Results

On the gender differences in chat messages and earnings across all three Douyu channels, *t*-test results showed that male streamers ($M = 10.69$, $SD = 1.57$) received more chat messages than female streamers ($M = 10.13$, $SD = 0.84$, $t(603) = 6.50$, $p < .001$). In contrast, female streamers ($M = 10.92$, $SD = 1.33$) earned more than male streamers ($M = 10.44$, $SD = 2.03$, $t(684) = -4.05$, $p < .001$).

H1 predicted that webcam use is associated with better live streaming outcomes. Results (see Model 1 in Table 2 and Model 1 in Table 3) showed that webcam use was positively associated with chat messages (coefficient = 0.36, $p < .001$), but was not significantly associated with earnings (coefficient = 0.18, $p = .057$), partially supporting H1.

H2 predicted that female streamers would be more likely to use webcams compared with male streamers. Results showed that 79.83% of female streamers turned on their webcams and only 27.34% of male streamers did so. This gender difference was statistically significant, $\chi^2$ (1, $N = 867$) = 238.42, $p < .001$, supporting H2.

For RQ1, as shown in Model 3 in Tables 2 and 3, the interaction between webcam use and gender was significant for both chat messages (coefficient = -0.64, $p < .001$) and earnings (coefficient = -0.47, $p < .01$). These results indicate that the positive associations between webcam use and both outcomes were weaker for female streamers than for male streamers.

H3 predicted that beauty filter use is associated with better live streaming outcomes. As shown in Model 2 in Tables 2 and 3, beauty filter use was positively associated with earnings (coefficient = 0.61, $p < .001$), but was negatively associated with chat messages (coefficient = -0.31, $p < .05$). Thus, H3 was partially supported.

H4 predicted that female streamers were more likely to use beauty filters than male streamers. Results showed that among webcam users, 99.18% of female streamers and 32.43% of male streamers used beauty filters, and the gender difference was significant ($\chi^2$ (1, $N = 479$) = 273.87, $p < .001$), supporting H4.

For RQ2, as shown in Model 4 in Tables 2 and 3, the interaction between gender and beauty filter use was not significant for chat messages (coefficient = -0.02, $p = .968$) or earnings (coefficient = 0.22, $p = .688$).

H5 predicted that greater title conformity to gender role expectations is associated with better live streaming outcomes. As shown in Model 1 in Tables 2 and 3, title conformity was positively associated with chat messages (coefficient = 0.10, $p < .05$), but not with earnings (coefficient = 0.03, $p = .601$), partially supporting H5.

H6 predicted that female and male streamers tended to customize their stream titles differently to conform to their gender role expectations. We compared male and female streamers' title content across the four identified themes and provided the two most representative keywords for each topic in Appendix 2. For the theme of *self-expression*, female streamers tended to highlight their emotional state with terms such as "不孤单 (Not alone)" and "啵叽 (Kiss sound)", while male streamers used phrases with fewer emotions like "加油 (Go for it)." Regarding the theme of *impression management*, female streamers emphasized their appearance and gentle nature using descriptors like "美女 (Beauty)" and "暖心 (Heartwarming)." In comparison, male streamers focused on their game skills and styles using terms such as "快准狠 (Fast, accurate, and ruthless)" and "操作秀 (Operation show)" in gaming channels, and their expertise using terms like "百科全书 (Encyclopedia)" in the Yanzhi channel. For the *interaction* theme, female streamers preferred to use tone particles or second person pronouns, such as "遇见你 (Meeting you)" and "来啦 (Here I cooome)", to close the distance with the viewers. In contrast, male streamers positioned themselves more as mentors and coaches using expressions like "教学局 (Teaching game)" and "带小姐姐 (Carry a girl)." Regarding the *stream content* theme, both male and female streamers in gaming channels emphasized their game characters, server details, and streaming schedules such as "微信区 (WeChat zone)" and "24 小时 (24 hours)." In Yanzhi channel, female streamers diversified their titles to include a variety of activities like "聊瓜 (Chat gossip)" and "唱歌 (Sing)", and male streamers rarely mentioned stream content in this channel. On the degree of conformity to gender role expectations, $t$-test results indicated that female streamers ($M$ = -2.64, $SD$ = 0.64) conformed significantly more than male streamers ($M$ = -2.80, $SD$ = 0.61, $t(859) = -3.71$, $p < .001$).

For RQ3, as shown in Model 5 in Tables 2 and 3, the interaction between gender and title conformity was not significant for chat messages (coefficient = -0.04, $p = .616$) or earnings (coefficient = -0.10, $p = .361$).

**Table 2**

*Regression Models Predicting Chat Messages*

| **Predictors** | **Model 1** | **Model 2** | **Model 3** | **Model 4** | **Model 5** |
|---|---|---|---|---|---|
| (Intercept) | 5.02*** | 6.40*** | 4.99*** | 6.40*** | 5.09*** |
| | (0.26) | (0.37) | (0.25) | (0.37) | (0.29) |
| Level | 0.01*** | 0.01*** | 0.01*** | 0.01*** | 0.01*** |
| | (0.00) | (0.00) | (0.00) | (0.00) | (0.00) |
| Duration | 0.00*** | 0.00*** | 0.00*** | 0.00*** | 0.00*** |
| | (0.00) | (0.00) | (0.00) | (0.00) | (0.00) |
| Viewer counts (log) | 0.39*** | 0.33*** | 0.39*** | 0.33*** | 0.39*** |
| | (0.02) | (0.04) | (0.02) | (0.04) | (0.03) |
| Female | -0.28*** | -0.42*** | 0.05 | -0.41 | -0.28*** |
| | (0.06) | (0.12) | (0.08) | (0.41) | (0.06) |
| Webcam on (H1) | 0.36*** | | 0.66*** | | 0.36*** |
| | (0.07) | | (0.08) | | (0.07) |
| Title conformity (H5) | 0.10* | 0.12* | 0.09* | 0.12* | 0.12* |
| | (0.04) | (0.05) | (0.04) | (0.05) | (0.06) |
| Filter on (H3) | | -0.31* | | -0.31* | |
| | | (0.14) | | (0.15) | |
| LOL | -0.03 | -0.11 | -0.12 | -0.11 | -0.03 |
| | (0.07) | (0.08) | (0.07) | (0.08) | (0.07) |
| HOK | 0.21** | 0.03 | 0.15 | 0.03 | 0.21** |
| | (0.08) | (0.10) | (0.08) | (0.10) | (0.08) |
| Gender × Webcam (RQ1) | | | -0.64*** | | |
| | | | (0.11) | | |
| Gender × Filter (RQ2) | | | | -0.02 | |
| | | | | (0.43) | |
| Gender × Title conformity (RQ3) | | | | | -0.04 |
| | | | | | (0.08) |
| Observations | 867 | 479 | 867 | 479 | 867 |
| $R^2$ | 0.70 | 0.71 | 0.71 | 0.71 | 0.70 |
| Adjusted $R^2$ | 0.70 | 0.70 | 0.71 | 0.70 | 0.70 |

*Note*. Standard errors are in parentheses; all *p* values are two-tailed. Female, webcam on, filter on, LOL, and HOK are dummy variables; male streamers, webcam off, filter off, and Yanzhi are the reference groups. Models 2 and 4 include webcam users only because beauty filters require webcam use. Chat messages and viewer counts were log-transformed. * $p < .05$. ** $p < .01$. *** $p < .001$.

**Table 3**

*Regression Models Predicting Earnings*

| Predictors | Model 1 | Model 2 | Model 3 | Model 4 | Model 5 |
|---|---|---|---|---|---|
| (Intercept) | 4.26*** | 4.66*** | 4.24*** | 4.67*** | 4.45*** |
| | (0.38) | (0.47) | (0.37) | (0.47) | (0.43) |
| Level | 0.04*** | 0.03*** | 0.04*** | 0.03*** | 0.04*** |
| | (0.00) | (0.00) | (0.00) | (0.00) | (0.00) |
| Duration | 0.00*** | 0.00*** | 0.00*** | 0.00*** | 0.00*** |
| | (0.00) | (0.00) | (0.00) | (0.00) | (0.00) |
| Viewer counts (log) | 0.38*** | 0.39*** | 0.38*** | 0.39*** | 0.37*** |
| | (0.04) | (0.05) | (0.04) | (0.05) | (0.04) |
| Female | 0.67*** | 0.02 | 0.91*** | -0.18 | 0.67*** |
| | (0.09) | (0.15) | (0.12) | (0.52) | (0.09) |
| Webcam on (H1) | 0.18 | | 0.40** | | 0.18 |
| | (0.10) | | (0.12) | | (0.10) |
| Title conformity (H5) | 0.03 | 0.03 | 0.02 | 0.03 | 0.09 |
| | (0.06) | (0.06) | (0.06) | (0.06) | (0.09) |
| Filter on (H3) | | 0.61*** | | 0.58** | |
| | | (0.18) | | (0.19) | |
| LOL | -0.79*** | -0.72*** | -0.86*** | -0.72*** | -0.80*** |
| | (0.10) | (0.10) | (0.10) | (0.10) | (0.10) |
| HOK | -0.35** | -0.51*** | -0.39*** | -0.51*** | -0.35** |
| | (0.11) | (0.12) | (0.11) | (0.12) | (0.11) |
| Gender × Webcam (RQ1) | | | -0.47** | | |
| | | | (0.17) | | |
| Gender × Filter (RQ2) | | | | 0.22 | |
| | | | | (0.55) | |
| Gender × Title conformity (RQ3) | | | | | -0.10 |
| | | | | | (0.11) |
| Observations | 867 | 479 | 867 | 479 | 867 |
| $R^2$ | 0.65 | 0.66 | 0.65 | 0.66 | 0.65 |

| Adjusted $R^2$ | 0.65 | 0.65 | 0.65 | 0.65 | 0.65 |
|---|---|---|---|---|---|

*Note*. Standard errors are in parentheses; all $p$ values are two-tailed. Female, webcam on, filter on, LOL, and HOK are dummy variables; male streamers, webcam off, filter off, and Yanzhi are the reference groups. Models 2 and 4 include webcam users only because beauty filters require webcam use. Earnings and viewer counts were log-transformed. * $p < .05$. ** $p < .01$. *** $p < .001$.

We further conducted robustness checks using channel-specific analyses (see Appendix 3, Tables 4A–6D). The channel-specific results were generally consistent with the direction of the pooled estimates, although several associations were not statistically significant within individual channels, likely reflecting reduced statistical power in smaller subsamples.

## Discussion

### *Differentiated Outcomes of CMC Tool Use*

Our findings do not support a uniformly positive association between CMC tool use and streaming outcomes. Webcam use and greater conformity of stream titles to gender role expectations were positively associated with chat messages but not with earnings. Beauty filter use, by contrast, showed a mixed pattern, being positively associated with earnings but negatively associated with chat messages. These findings suggest that different self-presentation tools can be associated with distinct, and sometimes opposing, outcomes. They also refine a straightforward application of the hyperpersonal model of CMC (Walther, 1996), indicating that optimized self-presentation does not necessarily generate both greater audience engagement and greater financial support.

The mixed association observed for beauty filter use is particularly noteworthy. Among webcam users, beauty filters may enhance perceived attractiveness and stimulate financial support from some viewers (Hou et al., 2020), while potentially reducing perceived authenticity or conversational immediacy (Tong et al., 2026), thereby discouraging active interaction. This pattern further suggests that earnings and chat messages capture different dimensions of audience response. Financial support may be driven by a relatively small group of high-contributing viewers, whereas chat messages reflect broader and more active participation (Wohn et al., 2018; Sjöblom & Hamari, 2017).

An alternative explanation involves self-selection or reverse causality. Streamers experiencing lower levels of chat engagement may be more likely to use filters as a compensatory strategy to enhance their visual appeal. Given the observational nature of the data, this result should be interpreted cautiously and replicated in future research.

***Under Societal Expectation: Female Streamers' Strategic Use of Webcams and Beauty Filters***

A central finding of this study is that female streamers were substantially more likely than male streamers to use webcams and beauty filters. As suggested by gender role theory (Eagly & Karau, 2002), women face stronger expectations to appear attractive, emotionally expressive, and relationally engaging, whereas men are more often evaluated on agentic qualities such as competence and performance. Because live streaming emphasizes visibility and audience interaction, platform tools may amplify these gendered expectations. CMC tools such as webcams and beauty filters are therefore not merely technical features; they also provide resources through which streamers respond to gendered standards of self-presentation.

Female streamers' greater use of visual tools illustrates how traditional gender norms persist within digital communication. At the same time, it demonstrates streamers' capacity to adapt strategically to the affordances of CMC (Walther, 1992). According to the hyperpersonal model of CMC (Walther, 1996), users can selectively manage available cues to optimize their self-presentation. In live streaming, streamers can curate their appearance, facial expressions, attire, and physical background to meet audience expectations and encourage support (Freeman & Wohn, 2020). Consistent with this argument, prior ethnographic research found that female streamers enlarged their webcam displays while minimizing gameplay screens (Zhang & Hjorth, 2019). White's (2006, p. 67) concept of the "webcam operator" similarly highlights the deliberate labor involved in women's strategic management of their visual presence. Thus, female streamers' greater use of webcams and beauty filters reflects both strategic agency and the constraints imposed by gendered audience expectations. Importantly, however, greater use of these visual tools should not be interpreted as evidence that female streamers necessarily receive greater benefits from them.

***Fulfilling but Limiting Effect of Webcam Use for Female Streamers***

The interaction models showed that gender moderated the association of webcam use with both chat messages and earnings. Webcam use was positively associated with both outcomes among male streamers, whereas neither association was significant among female streamers. This finding illustrates the persistence of gendered expectations in online environments (Postmes & Spears, 2002). Viewers may regard visual presence as a baseline expectation for female streamers but as an additional benefit when provided by male streamers. Consequently, webcam use may satisfy existing expectations for women without generating additional rewards, whereas it may exceed expectations and provide greater benefits for men.

This interpretation highlights the dual consequences of gender-role conformity. Although conformity may reduce the risk of negative evaluations or prejudice (Eagly & Karau, 2002), it may provide limited additional rewards when the behavior is already expected. Platforms and streamers should therefore encourage more diverse forms of self-presentation that extend beyond conventional gender norms, potentially fostering both greater creativity and a more inclusive streaming environment.

***Gender Dynamics in Stream Titles for Online Identity Development***

Prior research on live streaming platforms and gaming communities has shed light on gendered disparities in viewer interactions. Studies have shown that male streamers receive more gameplay-related comments, whereas female streamers are more frequently subjected to appearance-focused and objectifying comments (Nakandala et al., 2017; Ruberg et al., 2019). However, this line of scholarship has primarily focused on viewer behavior, neglecting the role of streamers in shaping their own online presence. Extending this work, our study demonstrates how streamers themselves actively construct gendered online identities through stream titles.

As revealed in our findings, female streamers more often emphasized emotions, personal experiences, vocal appeal, and relational connection. They also used second-person pronouns to reduce perceived distance and encourage viewer interaction. Male streamers, by contrast, more frequently highlighted gaming skills, competitive achievements, and expertise, positioning themselves as authorities or mentors. These patterns reflect broader gender role expectations and show that stream titles function not only as informational cues but also as tools for communicating gendered identities.

Furthermore, stream titles that follow familiar gendered conventions may create a sense of familiarity and comfort, potentially attracting viewers and fostering chat interactions, but this pattern may not translate into greater financial support. At the same time, such practices may reproduce stereotypes by repeatedly associating women with appearance and intimacy and men with competence and authority, which may inadvertently encourage some viewers to make inappropriate and even sexually harassing comments toward certain streamers, especially female streamers (Ruberg et al., 2019). Thus, although stream titles provide opportunities for strategic self-presentation, gendered expectations may also constrain the range of identities that streamers can effectively express.

### *Limitations and Future Directions*

Several limitations should be acknowledged. First, streamer gender was inferred by coders from visual and audio cues rather than obtained through self-reports. This approach may introduce perceptual bias and does not capture gender identities beyond binary categories. Second, the generalizability of our findings may be limited by our sampling method of choosing streamers only from the three most popular channels on Douyu. Future studies can diversify their samples to include less popular channels to achieve a comprehensive understanding of user behaviors on the platform. Third, this study only examined three CMC tools, namely the use of webcams, beauty filters, and stream titles. Analysis considering other CMC tools in live streaming, such as stream cover images and decoration of the live stream interface, would offer a more nuanced portrayal of the strategies streamers employ to engage their viewers effectively. Last, because the data are observational, the identified associations between CMC tools and streaming outcomes should not be interpreted as causal effects. Future research could use longitudinal or experimental designs to examine the causal consequences of CMC tool use.

### *Conclusion*

Previous research has largely overlooked streamers' use of CMC tools and gender differences in their usage. Drawing on gender role and CMC theories, this study shows that self-presentation tools have differentiated associations with streaming outcomes and that substantial gender disparities exist in their use. Webcam use and gender-conforming stream titles were positively associated with chat messages, whereas beauty-filter use was positively associated with earnings but negatively associated with chat messages. Female streamers relied more heavily on visual strategies such as webcams and beauty filters, yet webcam use benefited female streamers less than male streamers. Together, these findings reveal how traditional gender roles continue to shape streamer behavior and audience responses. Streamers and platforms should therefore reconsider how gendered expectations structure digital interaction and support a broader, fairer range of self-presentation practices.

## Appendices

### *Appendix 1. Identifying Active Streams*

To attract and retain more viewers, the Douyu algorithm rewards streamers who broadcast for longer durations and more frequently. Consequently, many streamers on Douyu often broadcast replays of their previous recordings while offline, creating the illusion that they are actually live streaming. While this approach may result in some viewer engagement, it does not generate real-time content or interactive responses from the streamer. To measure the actual broadcast duration of the streamer, we identified the active streams where the streamer was actually present. Therefore, in this study, we needed to identify the active streams to accurately measure the duration of streamers' streams. We address this challenge through approximating the volume of chat messages within specific time intervals. Specifically, we discretized continuous data from June 6 to July 31, 2022 into numerous consecutive ten-minute windows. Subsequently, we employed the following formula to assess the volume of chat messages within each window, thereby simulating the stream status:

$$O = \begin{cases} 0, if\ C < M - 0.25 \times S \\ 1, if\ C \geq M - 0.25 \times S \end{cases}$$

where the variable *C* refers to the volume of chat messages within a specific time window, *M* denotes the mean of chat messages across all windows, and *S* is the standard deviation of the chat messages across all windows. If *O* is 0, it indicates an inactive stream without a streamer during the corresponding time window; conversely, if *O* is 1, it suggests an active stream with a streamer.

*Appendix 2. Differences of Stream Title Customization between Male and Female Streamers*

| Theme | Operational definitions | Channel | Gender | Num. of topics | Top 2 words in each topic |
|---|---|---|---|---|---|
| Self-expression | A topic in which the high-frequency words present a streamer's personal states, emotions and greetings. | LOL | F | 2 | 不孤单 (Not alone), 晚安 (Good night); 啵叽 (Kiss sound), 陶冶情操 (Cultivate one's taste) |
| | | | M | 1 | 加油 (Go for it), 永恩 (Yone) |
| | | HOK | F | 2 | 来看看 (Come and see), 和你 (With you); 好起来了 (Getting better), 噜噜噜 (Lululu) |
| | | | M | 0 | / |
| | | Yanzhi | F | 5 | 有人疼 (Someone cares), 从头开始 (Start over); 陪伴 (Accompany), 相逢 (Meet); 翘首以盼 (Eagerly await), 重逢 (Reunion); 平安(Peace), 喜乐 (Joy); 开心 (Happy), 珍惜 (Cherish) |
| | | | M | 2 | 祝你好运 (Wish you good luck), 从头再来 (Start again); 有缘再见 (Fate to meet again), 快乐 (Happy) |
| Impression management | A topic in which the high-frequency words present a streamer's personal features that they want to deliver to viewers including personalities, appearances and vocal characters, and descriptions of their abilities. | LOL | F | 8 | 摄像头 (Webcam), 每天都在 (Stream everyday); 主播 (Streamer), 新主播 (New streamer); 小甜饼 (Sweetie), 你的小猫咪 (Your kitty); 心态好 (Good mindset), 暖心 (Heartwarming); 美好 (Wonderful), 美女 (Beauty); 脱颖而出 (Stand out), 御音 (Voice of a domineering lady); 最拉 (The worst) 很强 (Very strong); 自闭 (Introvert), 劳模 (Model worker) |
| | | | M | 7 | 天花板 (Ceiling), 教学 (Teaching); 国服 (National server), 第一 (First); 冲第一 (Rush for first), 峡谷之巅 (Kings Canyon's peak); 王者 (Champion), 金牌导师 (Gold medal mentor); ad 王 (AD king), MVP (Most |
| | | | | | Valuable Player); 上大师 (Reach master), 艾欧尼亚 (Ionia); 最强 (Strongest), 咔咔乱杀 (Random killing) |
| | | HOK | F | 6 | 早起 (Get up early), 专一 (Devoted); 夹子 (Valley girl accent), 人美声甜 (Beautiful and sweet-voiced); 国服 (National server), 上王者 (Reach champion); 新主播 (New streamer), 新赛季 (New season); 江西 (Jiangxi Province), 天花板 (Ceiling); 笨蛋 (Fool), 超稳 (Super stable) |
| | | | M | 8 | 连胜 (Winning streak), 上分如饮水 (Rank up easily); 全国第一 (National first), 稳国服 (Stably ranked on the national server leaderboard); 快准狠 (Fast, accurate, and ruthless), 操作秀 (Operation show); 我菜 (I'm unskilled), 努力 (Hardworking); 自信 (Confident), 开心 (Happy); 男孩 (Boy), 爱笑 (Loves to smile); 笨鸟 (Dumb bird), 懂吗 (Understand?); 超神 (Legendary), 斩杀 (Kill) |
| | | Yanzhi | F | 5 | 美少女 (Beautiful girl), 大眼 (Big eyes); 女猪播 (Female piggy streamer), 小棉袄 (Little cotton-padded jacket); 漂亮 (Beautiful), 开心 (Happy); 自由 (Freedom), 随性 (Casual); 拆迁户 (Demolition household), 微胖 (Slightly chubby) |
| | | | M | 1 | 保星 (Relegation), 百科全书 (Encyclopedia) |
| Interaction | A topic in which the high-frequency words show streamer's desire to interact with viewers. | LOL | F | 4 | 上分 (Rank up), 有车位 (Have a gamespot); 通宵车 (All-night gamespot), 俩妹妹 (Two beauties); 大乱斗 (Big brawl), 大胆点 (Be bolder); 见到你 (Seeing you), 遇见你 (Meeting you) |
| | | | M | 2 | 教学局 (Teaching game), 细节(Details); 开通钻粉 (Become diamond fans), 带小姐姐 (Carry a girl) |
| | | HOK | F | 2 | 上车啦 (Come to play), 上分 (Rank up); 来啦 (Here I cooome), 房间 (Room) |
| | | | M | 3 | 闯一闯 (Take a chance), 宇宙第一 (Universe's first); 冲前十 (Rush top ten), 前十 (Top ten); 等你 (Waiting for you), 创始人 (Founder) |
| | | Yanzhi | F | 2 | 看我 (Look at me), 等你 (Waiting for you); 哥哥 (Brother), 回来啦 (I'm back) |
| | | | M | 0 | / |
| Stream content | A topic in which the high-frequency words present stream contents and schedules. | LOL | F | 2 | 峡谷 (Kings Canyon), 千分 (Thousand points); 网一 (Bill Gewert), 发车 (Play) |
| | | | M | 3 | 休息 (Rest), 请假 (Take a leave); 大乱斗 (Big brawl), 网一 (Bill Gewert); 开船 (Make a bold move), 妖姬 (The Deceiver) |
| | | HOK | F | 1 | 双区 (WeChat and QQ zone), 微信区 (WeChat zone) |
| | | | M | 3 | 巅峰 (Peak), 24 小时 (24 hours); 每天直播 (Stream everyday), 全天 (All day); 靓仔 (Handsome), 鬼谷子 (Guiguzi: A hero in HOK) |
| | | Yanzhi | F | 2 | 唱歌 (Sing), 粤语 (Cantonese); 聊瓜 (Chat about gossips), 聊趣事 (Chat about fun things) |
| | | | M | 0 | / |

*Note.* F=Female; M= Male

***Appendix 3. Additional Analysis***

We conducted additional analyses to test our hypotheses separately within each of the three channels. Below, we report the results. The analyses include descriptive statistics, gender differences in webcam and beauty filter use, and the associations of webcam use, beauty-filter use, and title conformity with chat messages and earnings, as well as the interactions of these variables with gender, in the LOL (Tables 4A–4E), HOK (Tables 5A–5E), and Yanzhi channels (Tables 6A–6D).

***Streamers in LOL***

**Table 4A**

*Descriptive Statistics of Streamers in LOL*

| Variable | Total ($N$ = 339) | | | Male ($n$ = 188, 55.5%) | | | Female ($n$ = 151, 44.5%) | | |
|---|---|---|---|---|---|---|---|---|---|
| | Mean | Median | S.D. | Mean | Median | S.D. | Mean | Median | S.D. |
| Webcam | 0.54 | — | — | 0.39 | — | — | 0.72 | — | — |
| Beauty filter | 0.65 ($N$ = 183) | — | — | 0.16 ($n$ = 74) | — | — | 0.98 ($n$ = 109) | — | — |
| Title conformity | -2.65 | -2.58 | 0.60 | -2.69 | -2.65 | 0.63 | -2.59 | -2.47 | 0.56 |
| Streamer level | 50.05 | 45.00 | 19.58 | 53.09 | 48.50 | 20.79 | 46.26 | 43.00 | 17.30 |
| Duration | 280.94 | 266.33 | 140.77 | 311.01 | 305.58 | 147.21 | 243.49 | 220.17 | 122.84 |
| Chat messages | 222,214.48 | 22,855.00 | 834,327.51 | 331,350.01 | 34,264.00 | 991,715.21 | 86,337.13 | 17,378.00 | 555,825.15 |
| Earnings | 176,546.51 | 31,352.75 | 575,754.40 | 219,628.67 | 24,466.56 | 674,436.68 | 122,907.80 | 37,201.91 | 417,943.93 |
| Viewer counts | 95,727.26 | 42,297.00 | 121,501.68 | 127,905.95 | 98,882.50 | 127,959.03 | 55,663.74 | 16,118.00 | 99,696.05 |

*Note.* Webcam and beauty-filter use were coded 0 = no and 1 = yes; therefore, their means represent the proportions coded 1. Beauty-filter statistics were calculated only for webcam users.

**Table 4B**

*Frequency and Chi-Square Results for the Differences of Using Webcams between Male and Female Streamers in LOL (N=339)*

| Webcam setting | Male | | Female | | $\chi^2$ |
|---|---|---|---|---|---|
| | n | % | n | % | |
| off | 114 | 60.64 | 42 | 27.81 | 35.01*** |
| on | 74 | 39.36 | 109 | 72.19 | |

*Note*. *** $p < .001$.

**Table 4C**

*Frequency and Chi-Square Results for the Differences of Using Beauty Filters between Male and Female Streamers in LOL (N=183)*

| Beauty filter setting | Male | | Female | | $\chi^2$ |
|---|---|---|---|---|---|
| | n | % | n | % | |
| off | 62 | 83.78 | 2 | 1.83 | 126.58*** |
| on | 12 | 16.22 | 107 | 98.17 | |

*Note*. *** $p < .001$.

**Table 4D**

*Regression Models Predicting Chat Messages in LOL*

| **Predictors** | **Model 1** | **Model 2** | **Model 3** | **Model 4** | **Model 5** |
|---|---|---|---|---|---|
| (Intercept) | 4.91*** | 5.14*** | 4.79*** | 5.16*** | 4.97*** |
| | (0.45) | (0.72) | (0.45) | (0.72) | (0.47) |
| Level | 0.03*** | 0.03*** | 0.03*** | 0.03*** | 0.03*** |
| | (0.00) | (0.00) | (0.00) | (0.00) | (0.00) |
| Duration | 0.00*** | 0.00*** | 0.00*** | 0.00*** | 0.00*** |
| | (0.00) | (0.00) | (0.00) | (0.00) | (0.00) |
| Viewer counts (log) | 0.31*** | 0.34*** | 0.32*** | 0.34*** | 0.31*** |
| | (0.04) | (0.07) | (0.04) | (0.07) | (0.04) |
| Female | -0.27** | -0.23 | 0.09 | 0.03 | -0.27** |
| | (0.09) | (0.23) | (0.13) | (0.56) | (0.09) |
| Webcam on (H1) | 0.34*** | | 0.60*** | | 0.34*** |
| | (0.09) | | (0.11) | | (0.09) |
| Title conformity (H5) | 0.07 | 0.11 | 0.07 | 0.11 | 0.10 |
| | (0.07) | (0.11) | (0.06) | (0.11) | (0.08) |
| Filter on (H3) | | -0.29 | | -0.23 | |
| | | (0.23) | | (0.25) | |
| Gender × Webcam (RQ1) | | | -0.63*** | | |
| | | | (0.16) | | |
| Gender × Filter (RQ2) | | | | -0.30 | |
| | | | | (0.61) | |
| Gender × Title (RQ3) | | | | | -0.07 |
| | | | | | (0.13) |
| Observations | 339 | 183 | 339 | 183 | 339 |
| $R^2$ | 0.79 | 0.80 | 0.79 | 0.80 | 0.79 |
| Adjusted $R^2$ | 0.78 | 0.80 | 0.79 | 0.80 | 0.78 |

*Note*. Standard errors are in parentheses. All *p* values in this table are two-tailed. Female, webcam on and filter on are dummy variables (male streamers, webcam off and filter off are the comparison group). Model 2 and Model 4 analyze streamers who have turned on their webcams, as beauty filters can only be applied when the webcam is on. Chat messages and viewer counts were log-transformed to adjust for skewness. * $p < .05$. ** $p < .01$. *** $p < .001$.

**Table 4E**

*Regression Models Predicting Earnings in LOL*

| Predictors | Model 1 | Model 2 | Model 3 | Model 4 | Model 5 |
|---|---|---|---|---|---|
| (Intercept) | 4.52*** | 3.41*** | 4.48*** | 3.29*** | 4.64*** |
| | (0.59) | (0.80) | (0.60) | (0.79) | (0.62) |
| Level | 0.05*** | 0.04*** | 0.05*** | 0.04*** | 0.05*** |
| | (0.00) | (0.00) | (0.00) | (0.00) | (0.00) |
| Duration | 0.00*** | 0.00** | 0.00*** | 0.00** | 0.00*** |
| | (0.00) | (0.00) | (0.00) | (0.00) | (0.00) |
| Viewer counts (log) | 0.24*** | 0.39*** | 0.25*** | 0.40*** | 0.24*** |
| | (0.06) | (0.07) | (0.06) | (0.07) | (0.06) |
| Female | 0.91*** | 0.76** | 1.02*** | -0.56 | 0.92*** |
| | (0.12) | (0.26) | (0.18) | (0.60) | (0.12) |
| Webcam on (H1) | 0.20 | | 0.28 | | 0.20 |
| | (0.11) | | (0.15) | | (0.11) |
| Title conformity (H5) | 0.09 | 0.06 | 0.09 | 0.05 | 0.14 |
| | (0.09) | (0.12) | (0.09) | (0.12) | (0.11) |
| Filter on (H3) | | 0.20 | | -0.06 | |
| | | (0.25) | | (0.27) | |
| Gender × Webcam (RQ1) | | | -0.19 | | |
| | | | (0.22) | | |
| Gender × Filter (RQ2) | | | | 1.59* | |
| | | | | (0.66) | |
| Gender × Title (RQ3) | | | | | -0.13 |
| | | | | | (0.17) |
| Observations | 339 | 183 | 339 | 183 | 339 |
| $R^2$ | 0.70 | 0.74 | 0.70 | 0.75 | 0.70 |
| Adjusted $R^2$ | 0.69 | 0.73 | 0.69 | 0.74 | 0.69 |

*Note*. Standard errors are in parentheses. All $p$ values in this table are two-tailed. Female, webcam on and filter on are dummy variables (male streamers, webcam off and filter off are the comparison group). Model 2 and Model 4 analyze streamers who have turned on their webcams, as beauty filters can only be applied when the webcam is on. Earnings and viewer counts were log-transformed to adjust for skewness. * $p < .05$. ** $p < .01$. *** $p < .001$.

***Streamers in HOK***

**Table 5A**

*Descriptive Statistics of Streamers in HOK*

| Variable | Total ($N$ = 301) | | | Male ($n$ = 196, 65.1%) | | | Female ($n$ = 105, 34.9%) | | |
|---|---|---|---|---|---|---|---|---|---|
| | Mean | Median | S.D. | Mean | Median | S.D. | Mean | Median | S.D. |
| Webcam | 0.23 | — | — | 0.08 | — | — | 0.51 | — | — |
| Beauty filter | 0.87 ($N$ = 69) | — | — | 0.47 ($n$ = 15) | — | — | 0.98 ($n$ = 54) | — | — |
| Title conformity | -2.84 | -2.81 | 0.59 | -2.92 | -2.93 | 0.57 | -2.70 | -2.63 | 0.62 |
| Streamer level | 47.80 | 45.00 | 18.39 | 47.44 | 46.00 | 19.17 | 48.47 | 45.00 | 16.90 |
| Duration | 290.09 | 274.17 | 176.17 | 298.91 | 281.00 | 195.61 | 273.63 | 259.83 | 131.79 |
| Chat messages | 136,421.88 | 24,208.00 | 642,885.31 | 188,507.48 | 26,171.00 | 791,391.13 | 39,195.44 | 23,754.00 | 56,981.57 |
| Earnings | 150,470.65 | 40,561.58 | 542,086.24 | 181,437.59 | 37,316.79 | 663,192.64 | 92,665.69 | 47,997.83 | 133,469.60 |
| Viewer counts | 53,103.02 | 18,434.00 | 82,088.87 | 65,254.81 | 20,202.00 | 93,774.30 | 30,419.68 | 15,954.00 | 46,394.07 |

*Note.* Webcam and beauty-filter use were coded 0 = no and 1 = yes; therefore, their means represent the proportions coded 1. Beauty-filter statistics were calculated only for webcam users.

**Table 5B**

*Frequency and Chi-Square Results for the Differences of Using Webcams between Male and Female Streamers in HOK (N=301)*

| Webcam setting | Male | | Female | | $\chi^2$ |
|---|---|---|---|---|---|
| | n | % | n | % | |
| off | 181 | 92.35 | 51 | 48.57 | 71.70*** |
| on | 15 | 7.65 | 54 | 51.43 | |

*Note*. *** $p < .001$.

**Table 5C**

*Frequency and Chi-Square Results for the Differences of Using Beauty Filters between Male and Female Streamers in HOK (N=69)*

| Beauty filter setting | Male | | Female | | $\chi^2$ |
|---|---|---|---|---|---|
| | n | % | n | % | |
| off | 8 | 53.33 | 1 | 1.85 | 23.08*** |
| on | 7 | 46.67 | 53 | 98.15 | |

*Note*. *** $p < .001$.

**Table 5D**

*Regression Models Predicting Chat Messages in HOK*

| **Predictors** | **Model 1** | **Model 2** | **Model 3** | **Model 4** | **Model 5** |
|---|---|---|---|---|---|
| (Intercept) | 4.47*** | 6.99*** | 4.54*** | 7.06*** | 4.43*** |
| | (0.41) | (0.87) | (0.40) | (0.89) | (0.46) |
| Level | 0.01*** | 0.01* | 0.01*** | 0.01* | 0.01*** |
| | (0.00) | (0.01) | (0.00) | (0.01) | (0.00) |
| Duration | 0.00*** | 0.00*** | 0.00*** | 0.00*** | 0.00*** |
| | (0.00) | (0.00) | (0.00) | (0.00) | (0.00) |
| Viewer counts (log) | 0.51*** | 0.38*** | 0.48*** | 0.37*** | 0.51*** |
| | (0.04) | (0.09) | (0.04) | (0.09) | (0.04) |
| Female | -0.19 | -0.94*** | 0.01 | -1.22 | -0.19 |
| | (0.10) | (0.26) | (0.11) | (0.70) | (0.10) |
| Webcam on (H1) | 0.16 | | 0.76*** | | 0.16 |
| | (0.11) | | (0.19) | | (0.11) |
| Title conformity (H5) | 0.14 | 0.35** | 0.12 | 0.35** | 0.13 |
| | (0.07) | (0.13) | (0.07) | (0.13) | (0.09) |
| Filter on (H3) | | -0.17 | | -0.24 | |
| | | (0.30) | | (0.34) | |
| Gender × Webcam (RQ1) | | | -0.91*** | | |
| | | | (0.23) | | |
| Gender × Filter (RQ2) | | | | 0.31 | |
| | | | | (0.74) | |
| Gender × Title (RQ3) | | | | | 0.02 |
| | | | | | (0.15) |
| Observations | 301 | 69 | 301 | 69 | 301 |
| $R^2$ | 0.72 | 0.80 | 0.74 | 0.80 | 0.72 |
| Adjusted $R^2$ | 0.72 | 0.78 | 0.73 | 0.78 | 0.72 |

*Note*. Standard errors are in parentheses. All *p* values in this table are two-tailed. Female, webcam on and filter on are dummy variables (male streamers, webcam off and filter off are the comparison group). Model 2 and Model 4 analyze streamers who have turned on their webcams, as beauty filters can only be applied when the webcam is on. Chat messages and viewer counts were log-transformed to adjust for skewness. * $p < .05$. ** $p < .01$. *** $p < .001$.

**Table 5E**

*Regression Models Predicting Earnings in HOK*

| **Predictors** | **Model 1** | **Model 2** | **Model 3** | **Model 4** | **Model 5** |
|---|---|---|---|---|---|
| (Intercept) | 1.54* | 0.85 | 1.57* | 0.63 | 1.56* |
| | (0.65) | (1.08) | (0.65) | (1.09) | (0.73) |
| Level | 0.04*** | 0.03*** | 0.04*** | 0.03*** | 0.04*** |
| | (0.00) | (0.01) | (0.00) | (0.01) | (0.00) |
| Duration | 0.00*** | 0.00 | 0.00*** | 0.00 | 0.00*** |
| | (0.00) | (0.00) | (0.00) | (0.00) | (0.00) |
| Viewer counts (log) | 0.60*** | 0.76*** | 0.59*** | 0.77*** | 0.60*** |
| | (0.06) | (0.11) | (0.06) | (0.11) | (0.06) |
| Female | 0.74*** | 0.43 | 0.83*** | 1.41 | 0.74*** |
| | (0.16) | (0.32) | (0.18) | (0.86) | (0.16) |
| Webcam on (H1) | -0.07 | | 0.21 | | -0.07 |
| | (0.18) | | (0.31) | | (0.18) |
| Title conformity (H5) | -0.10 | 0.02 | -0.11 | 0.01 | -0.09 |
| | (0.11) | (0.15) | (0.11) | (0.15) | (0.15) |
| Filter on (H3) | | 0.40 | | 0.64 | |
| | | (0.37) | | (0.42) | |
| Gender × Webcam (RQ1) | | | -0.42 | | |
| | | | (0.38) | | |
| Gender × Filter (RQ2) | | | | -1.11 | |
| | | | | (0.90) | |
| Gender × Title (RQ3) | | | | | -0.01 |
| | | | | | (0.23) |
| Observations | 301 | 69 | 301 | 69 | 301 |
| $R^2$ | 0.67 | 0.80 | 0.67 | 0.80 | 0.67 |
| Adjusted $R^2$ | 0.66 | 0.78 | 0.66 | 0.78 | 0.66 |

*Note*. Standard errors are in parentheses. All $p$ values in this table are two-tailed. Female, webcam on and filter on are dummy variables (male streamers, webcam off and filter off are the comparison group). Model 2 and Model 4 analyze streamers who have turned on their webcams, as beauty filters can only be applied when the webcam is on. Earnings and viewer counts were log-transformed to adjust for skewness. * $p < .05$. ** $p < .01$. *** $p < .001$.

***Streamers in Yanzhi***

**Table 6A**

*Descriptive Statistics of Streamers in Yanzhi*

| Variable | Total (*N* = 227) | | | Male (*n* = 22, 9.7%) | | | Female (*n* = 205, 90.3%) | | |
|---|---|---|---|---|---|---|---|---|---|
| | Mean | Median | S.D. | Mean | Median | S.D. | Mean | Median | S.D. |
| Webcam | 1.00 | — | — | 1.00 | — | — | 1.00 | — | — |
| Beauty filter | 0.98 | — | — | 0.77 | — | — | 1.00 | — | — |
| Title conformity | -2.63 | -2.49 | 0.71 | -2.57 | -2.55 | 0.65 | -2.64 | -2.47 | 0.71 |
| Streamer level | 50.89 | 48.00 | 19.71 | 64.45 | 62.50 | 18.41 | 49.44 | 47.00 | 19.33 |
| Duration | 262.01 | 259.67 | 102.37 | 311.95 | 300.67 | 106.38 | 256.65 | 256.17 | 100.73 |
| Chat messages | 38,823.30 | 29,494.00 | 48,780.22 | 67,528.95 | 30,823.00 | 83,535.18 | 35,742.70 | 28,851.00 | 42,654.27 |
| Earnings | 188,021.35 | 93,057.06 | 348,831.83 | 358,109.67 | 197,636.13 | 370,366.04 | 169,767.96 | 83,150.26 | 342,390.30 |
| Viewer counts | 59,846.59 | 20,582.00 | 153,145.01 | 59,509.55 | 16,593.50 | 128,375.76 | 59,882.77 | 20,957.00 | 155,840.18 |

*Note.* Webcam and beauty-filter use were coded 0 = no and 1 = yes; therefore, their means represent the proportions coded 1. Beauty-filter statistics were calculated only for webcam users.

**Table 6B**

*Fisher's Exact Test Results for the Differences of Using Beauty Filters between Male and Female Streamers in Yanzhi (N=227)*

| Beauty filter setting | Male | | Female | | *p* |
|---|---|---|---|---|---|
| | *n* | % | *n* | % | |
| off | 5 | 22.73 | 0 | 0.00 | $p < .001$ |
| on | 17 | 77.27 | 205 | 100.00 | |

**Table 6C**

*Regression Models Predicting Chat Messages in Yanzhi*

| **Predictors** | **Model 1** | **Model 2** |
|---|---|---|
| (Intercept) | 7.67*** | 8.32*** |
| | (0.34) | (0.53) |
| Level | -0.00 | -0.00 |
| | (0.00) | (0.00) |
| Duration | 0.00*** | 0.00*** |
| | (0.00) | (0.00) |
| Viewer counts (log) | 0.22*** | 0.22*** |
| | (0.03) | (0.03) |
| Female | -0.26* | -0.22* |
| | (0.11) | (0.11) |
| Filter on (H3) | -0.08 | -0.15 |
| | (0.21) | (0.21) |
| Title conformity (H5) | 0.06 | 0.27 |
| | (0.04) | (0.14) |
| Gender × Title (RQ3) | | -0.23 |
| | | (0.15) |
| Observations | 227 | 227 |
| $R^2$ | 0.62 | 0.62 |
| Adjusted $R^2$ | 0.61 | 0.61 |

*Note*. All streamers in this channel turned on the webcam and all female streamers used beauty filters; Female and filter on are dummy variables (male streamers and filter off are the comparison group). Chat messages and viewer counts were log-transformed to adjust for skewness. * $p < .05$. ** $p < .01$. *** $p < .001$.

**Table 6D**

*Regression Models Predicting Earnings in Yanzhi*

| Predictors | Model 1 | Model 2 |
|---|---|---|
| (Intercept) | 6.45*** | 6.83*** |
| | (0.70) | (1.10) |
| Level | 0.02*** | 0.02*** |
| | (0.00) | (0.00) |
| Duration | 0.00*** | 0.00*** |
| | (0.00) | (0.00) |
| Viewer counts (log) | 0.31*** | 0.31*** |
| | (0.07) | (0.07) |
| Female | -0.55* | -0.53* |
| | (0.22) | (0.23) |
| Filter on (H3) | 0.25 | 0.21 |
| | (0.43) | (0.44) |
| Title conformity (H5) | -0.01 | 0.12 |
| | (0.08) | (0.29) |
| Gender × Title (RQ3) | | -0.14 |
| | | (0.31) |
| Observations | 227 | 227 |
| $R^2$ | 0.55 | 0.55 |
| Adjusted $R^2$ | 0.54 | 0.54 |

*Note*. All streamers in this channel turned on the webcam and all female streamers used beauty filters; Female and filter on are dummy variables (male streamers and filter off are the comparison group). Earnings and viewer counts were log-transformed to adjust for skewness.

* $p < .05$. ** $p < .01$. *** $p < .001$.